\documentclass[%
reprint,
superscriptaddress,
citeautoscript,
amsmath,amssymb,
aps,
prl,
]{revtex4-1}

\usepackage{graphicx}
\usepackage{dcolumn}
\usepackage{bm}
\usepackage{color}
\usepackage{subfigure}
\usepackage{hyperref}
\usepackage[mathlines]{lineno}

\usepackage[dvipsnames]{xcolor}

\begin{document}


\title{Hot-electron mediated ion diffusion in proton-irradiated magnesium oxide}

\author{Cheng-Wei Lee}
\affiliation{Department of Materials Science and Engineering, University of Illinois at Urbana-Champaign, Urbana, IL 61801, USA}

\author{Andr\'e Schleife}
\email{schleife@illinois.edu}
\affiliation{Department of Materials Science and Engineering, University of Illinois at Urbana-Champaign, Urbana, IL 61801, USA}
\affiliation{Frederick Seitz Materials Research Laboratory, University of Illinois at Urbana-Champaign, Urbana, IL 61801, USA}
\affiliation{National Center for Supercomputing Applications, University of Illinois at Urbana-Champaign, Urbana, IL 61801, USA}

\date{\today}

\begin{abstract}
Highly energetic ions that impact materials have applications from semiconductor industry to medicine, and are fundamentally interesting as they trigger multi-length and time-scale processes.
In particular, they excite electrons into non-thermalized energy distributions with subsequent non-equilibrium electron-electron and electron-ion dynamics.
In order to achieve a quantitative description of these, we propose a general first-principles framework that bridges time scales from ultrafast electron dynamics directly after impact, to ion diffusion over migration barriers in semiconductors.
We apply it to magnesium oxide under proton irradiation and discover a diffusion mechanism that is mediated by hot electrons.
Our quantitative simulations show that this mechanism strongly depends on the projectile-ion velocity.
This indicates that it may occur only at a specific penetration depth in the target and that it can be triggered by varying the kinetic energy of the particle radiation.
Either of these predictions should facilitate direct experimental observation of this effect and significantly advances current understanding of non-equilibrium electron-ion dynamics.
\end{abstract}

\maketitle

Energetic charged-particle radiation has exciting applications including modern research, semiconductor industry, and medicine:
Helium-ion microscopy shows excellent resolution \cite{Notte:2006} and is superior to traditional electron microscopy, e.g.\ for insulating systems and bio-materials \cite{Fitzpatrick:2013}.
Focused-ion-beam techniques achieve micro- and nano-scale structuring for photonic, plasmonic, and microelectromechanical systems \cite{Kim:2012,Lindquist:2012,Krasheninnikov:2010}.
Charged-ion therapy is becoming a competitive alternative to X-ray treatment because of better spatial control of energy deposition in the human body, reducing side effects \cite{Durante:2010,Schardt:2010}.
These successes rely on quantitative understanding of fundamental interactions between particle radiation and target materials.

Highly energetic ions, carrying keV or MeV of kinetic energy, trigger multi-length- and time-scale processes, depending on mass, charge, and kinetic energy of the projectile ion, the impact parameter of the scattering event, and the target material \cite{Correa:2018,Wirth:2004}.
Generally, the underlying scattering physics divides the interaction between charged particles and target into two regimes:
Fast, charged particles, typically at early stages of the interaction, scatter inelastically.
In this electronic-stopping regime, where ions are too slow to respond, kinetic energy of particle radiation translates to hot, excited carriers.
This manifests itself as electronic friction, slowing down the projectile and rendering elastic scattering with lattice ions more likely, and eventually dominating.
The interaction then becomes similar to that for non-ionizing particle radiation, which has been well studied in experiment, theory, and computation \cite{Averback:1997,Wirth:2004,Raine:2017}.

Since high-energy projectiles significantly drive the electronic system of the target out of equilibrium, they are an ideal \emph{probe} of hard to access, ultrafast non-equilibrium electron-ion physics \cite{Susi:2014}.
Immediately after impact, the electronic system is in a highly excited, \emph{non-thermalized} state.
Subsequent thermalization towards a Fermi distribution with a well-defined temperature takes tens to hundreds of femtoseconds \cite{Nie:2014,Bernadi:2014,Jhalani:2017,Richter:2017,Harutyunyan:2015}, depending on (electron-electron and electron-phonon) scattering mechanisms and the target (semiconductor vs.\ metal).
After thermalization, hot electrons cool over tens of picoseconds by equilibration with the lattice \cite{White:2014,Lin:2017,Elsayed-Ali:1993,Sadasivam:2017}.
Even though ion dynamics occurs on a comparable time scale of hundreds of femtoseconds, quantified by attempt frequencies of about 13 THz \cite{Runevall:2011}, it is not well understood whether non-thermalized excited carriers and thermalized hot carriers affect atomic \emph{diffusion}.

Hence, ultrafast interactions between hot carriers and lattice ions are subject of intensive research.
While ionizing radiation is known to enhance diffusion of point defects in non-metals in the regime of large electronic stopping \cite{Zinkle:1995,Zinkle:2018,Zhang:2015}, underlying mechanisms are poorly explored.
It is proposed, supported by first-principles calculations \cite{Lei:2015,Mulroue:2011}, that a change of the defect charge state significantly influences diffusion \cite{zinkle:1996,Bourgoin:1989}.
However, it remains unclear how to obtain the charge state for a specific projectile and kinetic energy.
Developing quantitative understanding of how hot, thermalized and non-thermalized electrons affect atomic diffusion requires an extension of the charged-defect picture.

\begin{figure}
\includegraphics[width=0.95\columnwidth]{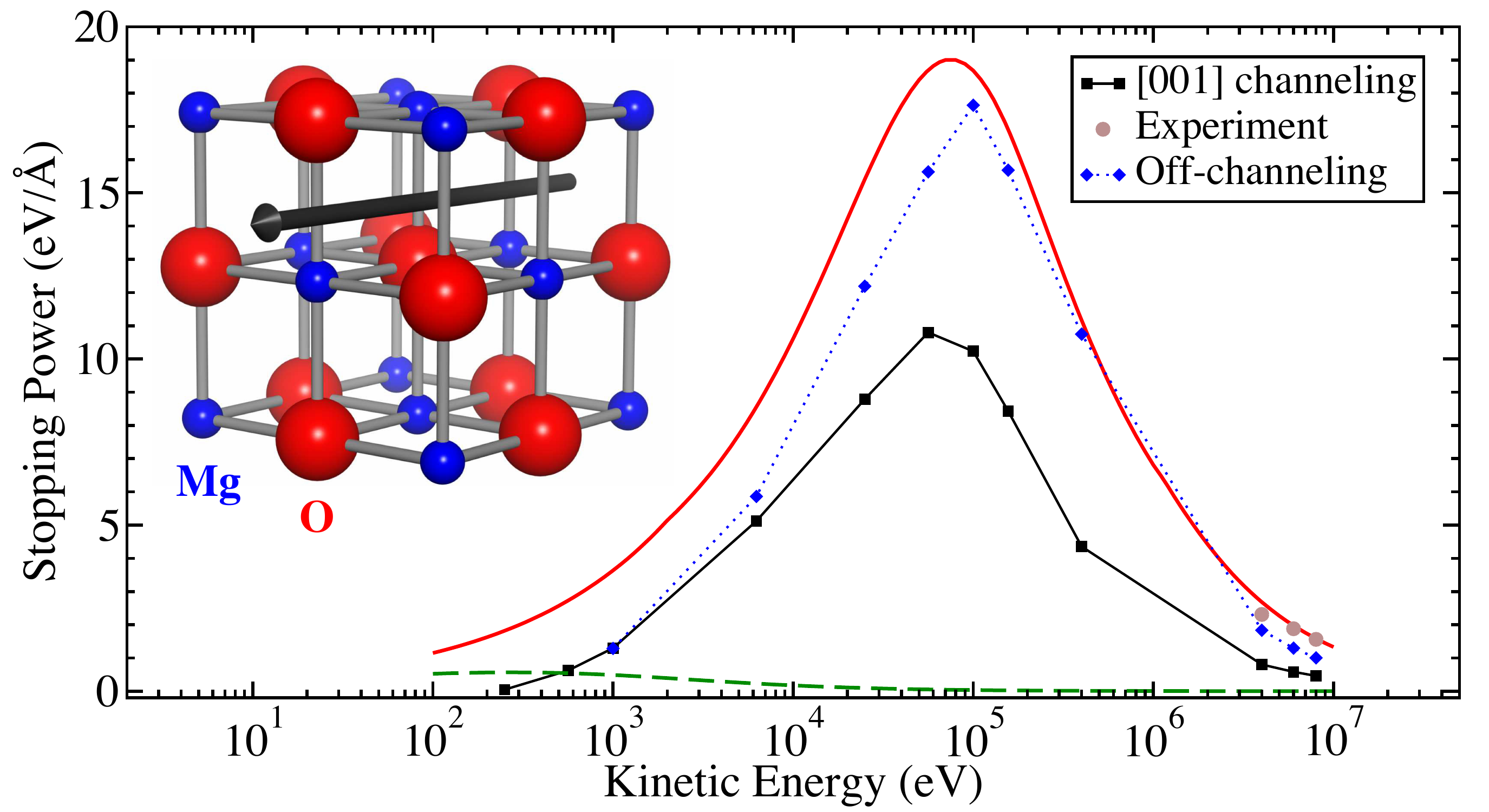}
\caption{\label{fig:vs_SRIM}(Color online.)
Electronic stopping of protons in rocksalt MgO.
RT-TDDFT results are shown for off-channeling (blue diamonds) and [001] channel (black squares), indicated as black arrow in the inset.
SRIM \cite{Ziegler:2010} simulations for a density of 3.60 g/cm$^3$ are shown for electronic (red solid) and nuclear (green dashed) stopping. 
Brown circles at high energy represent the only experiment \cite{Clark:1970} for off-channeling protons in MgO included in SRIM.
}
\end{figure}

We devise a first-principles framework that describes electronic response to irradiation and electron-ion dynamics ensuing after the initial excitation, to explore the impact of hot electrons on point-defect diffusivity.
We study oxygen vacancies in magnesium oxide (MgO, Fig.\ \ref{fig:vs_SRIM}) as test case for which extensive first-principles studies \cite{Alfe:2005,Rinke:2012,Runevall:2011,Ertekin:2013,Lin:2013} showed that density functional theory provides qualitatively correct defect levels.
Since a full quantum-mechanical treatment of coupled electron-ion dynamics in solids is computationally infeasible, 
we employ cutting-edge approximations:
Nuclei are described by classical Coulomb potentials, which is justified by large ion masses of the target material studied here.
Initially, the projectile moves too fast through the target for ions to respond and we explicitly simulate creation of electronic excitations.
To compute these time-dependent hot-electron distributions we use real-time time-dependent density functional theory (RT-TDDFT) \cite{Runge:1984} in Ehrenfest molecular dynamics (EMD) simulations \cite{Ehrenfest:1927,Marx:2009,Schleife:2015}.
Electrons are excited at a rate that corresponds to transfer of kinetic energy of the projectile to the electronic system, i.e.\ electronic stopping $S$,
\begin{equation}
\label{eq:stopping}
S = dE/dx.
\end{equation}
RT-TDDFT has been demonstrated to accurately describe electronic stopping in metals \cite{Correa:2012,Zeb:2012,Bi:2017,Schleife:2014,Schleife:2015,Quashie:2016}, semiconductors \cite{Hatcher:2008,Ullah:2015,Lim:2016,Yost:2016,Yost:2017,Lee:2018}, insulators \cite{Pruneda:2007,Li:2017}, nanostructures \cite{Krasheninnikov:2007,Quijada:2007,Ojanpera:2014,Wang:2015}, water \cite{Reeves:2016,Reeves:2017}, and warm dense matter \cite{Magyar:2016}.

\begin{figure}
\includegraphics[width=0.95\columnwidth]{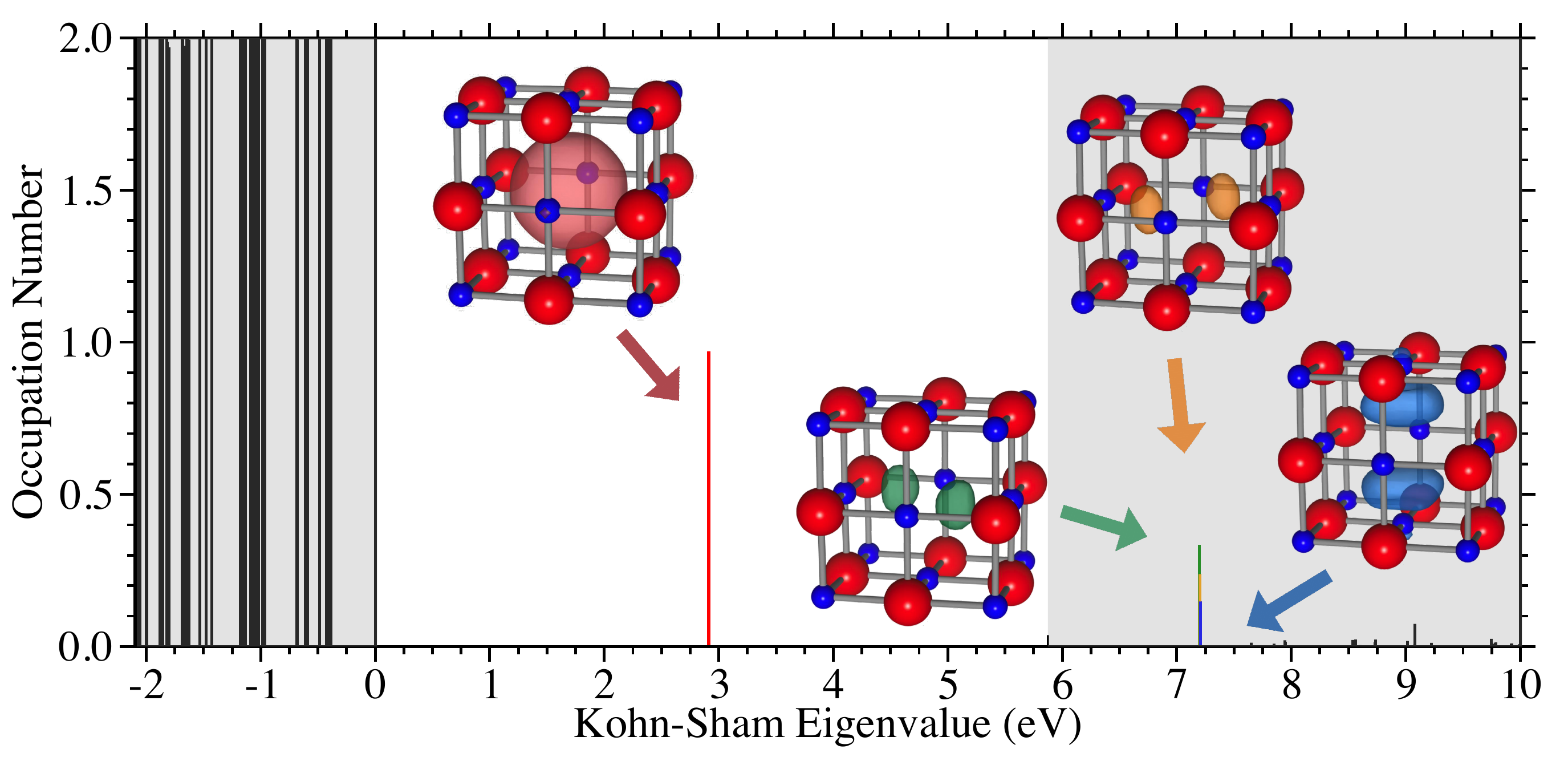}
\caption{\label{fig:change_occ}(Color online.)
Vertical lines indicate occupation numbers of adiabatic KS states near the band gap (unshaded area) after proton irradiation ($v$=0.15 at.\ u.).
Valence-band maximum is used as energy zero.
Isosurfaces of partial charge density at 50\,\% of the maximum are shown as insets for mid-gap state (red) and first three excited localized states (green, orange, and blue, respectively).
}
\end{figure}

Explicit simulations rely on the Qb@ll \cite{Schleife:2014,qball:2017} and VASP codes \cite{Kresse:1999,Kresse:1996} to perform ground-state DFT calculations for MgO without (216-atom cell, Fig.\ \ref{fig:vs_SRIM}) and with oxygen vacancy (215-atom cell, Fig.\ \ref{fig:change_occ}) with computational parameters described in Ref.\ \onlinecite{cutoffs}.
Time-dependent excited electronic states are computed using EMD for a proton moving along the center of the [001] channel closest to the oxygen vacancy, using Qb@ll \cite{Schleife:2014}, see computational details in Ref.\ \onlinecite{tdsimulations}.

\begin{figure}
\includegraphics[width=0.98\columnwidth]{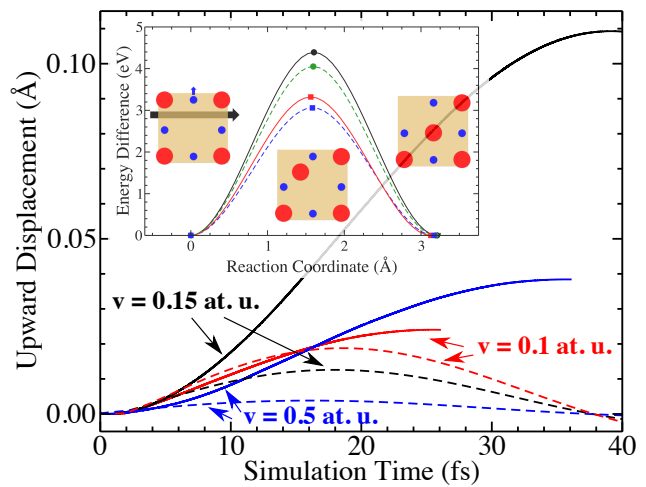}
\caption{\label{fig:Mg2_111_disp}(Color online.)
Displacement of a Mg atom (indicated by blue arrow in inset) for projectile velocities of 0.1 at.\ u.\ (red), 0.15 at.\ u.\ (black), and 0.5 at.\ u.\ (blue).
Solid and dashed lines are for EMD and Born-Oppenheimer MD, respectively.
Inset shows a diffusion path and its migration barrier computed using ground-state DFT (black circle and solid line), effective charge state (red square and solid line), Mermin DFT (green circle and dashed line), and c-DFT (blue square and dashed line) to approximate the excited state resulting from proton irradiation ($v$=0.15 at.\ u.).
}
\end{figure}

After the projectile traverses the simulation cell, during which electrons are excited, it is removed and we continue EMD simulations to reveal subsequent electron-ion dynamics.
After tens of femtoseconds, depending on the velocity of the incident proton, the displacement of Mg atoms adjacent to the vacancy reaches its maximum (after 40 fs for the atom shown in inset of Fig.\ \ref{fig:Mg2_111_disp}).
Since the high computational cost of EMD simulations prevents longer runs, we instead compute time-dependent occupation numbers of single-particle Kohn-Sham (KS) states via
\begin{equation}
\label{eq:occ}
f_{i}(t)=\sum_{j=1}\left|\left\langle\phi_{i}|\psi_{j}(t)\right\rangle\right|^{2}
\end{equation}
where $\phi_{i}$ are adiabatic KS ground-state orbitals of the instantaneous atomic configuration and $\psi_{j}(t)$ are non-adiabatic time-dependent states.
We then account for the statistical nature and longer time scale of diffusion using transition-state theory and compute migration barrier and defect diffusivity in the presence of hot carriers, using these occupations as constraint in the nudged-elastic band (NEB) method \cite{Henkelman:2000_a,Henkelman:2000_b}.

Figure \ref{fig:vs_SRIM} shows that our RT-TDDFT results for electronic stopping in ideal bulk MgO agree well with the ``The Stopping and Range of Ions in Matter'' (SRIM) \cite{Ziegler:2010} Monte Carlo package, parameterized using experimental input.
While for channeling protons we observe an underestimation of electronic stopping across the entire kinetic-energy range, using an off-channeling trajectory gives rise to much better agreement between RT-TDDFT and SRIM near and past the electronic-stopping maximum.
We attribute this to the reduced ability of channeling projectiles to excite semi-core electrons concentrated mostly near atomic positions \cite{Schleife:2015,Yost:2017}.
However, Fig.\ \ref{fig:vs_SRIM} also shows a remaining deviation at low proton kinetic energies.
This can be explained by the approximate character of SRIM, which averages stopping of Mg and O atoms to obtain stopping for compound MgO.
While this approximation is valid at high kinetic energy, it overestimates stopping at low kinetic energy \cite{Feng:1974, Langley:1976, Abril:2002} where band-structure effects become important in compounds.

Next, we analyze how a neutral oxygen vacancy affects electronic stopping and identified two competing mechanisms, depending on the projectile kinetic energy.
For slow projectiles, transitions between localized vacancy states (see Fig.\ \ref{fig:change_occ}) increase electronic stopping;
the enhancement peaks for a proton velocity of $v$=0.15 at.\ u.\ ($\approx$ 0.56 keV).
Contrary, fast projectiles predominantly interact with the local charge density \cite{Winter:2003,Ullah:2015} that is decreased near the vacancy, compared to a perfect crystal.
This effect is more important the higher the projectile velocity, and for more than $v$=1.0 at.\ u., a net reduction of electronic stopping is observed, compared to ideal bulk MgO (see supplemental material at [URL] for velocity-dependent local stopping with and without vacancy).

\begin{figure}
\includegraphics[width=0.98\columnwidth]{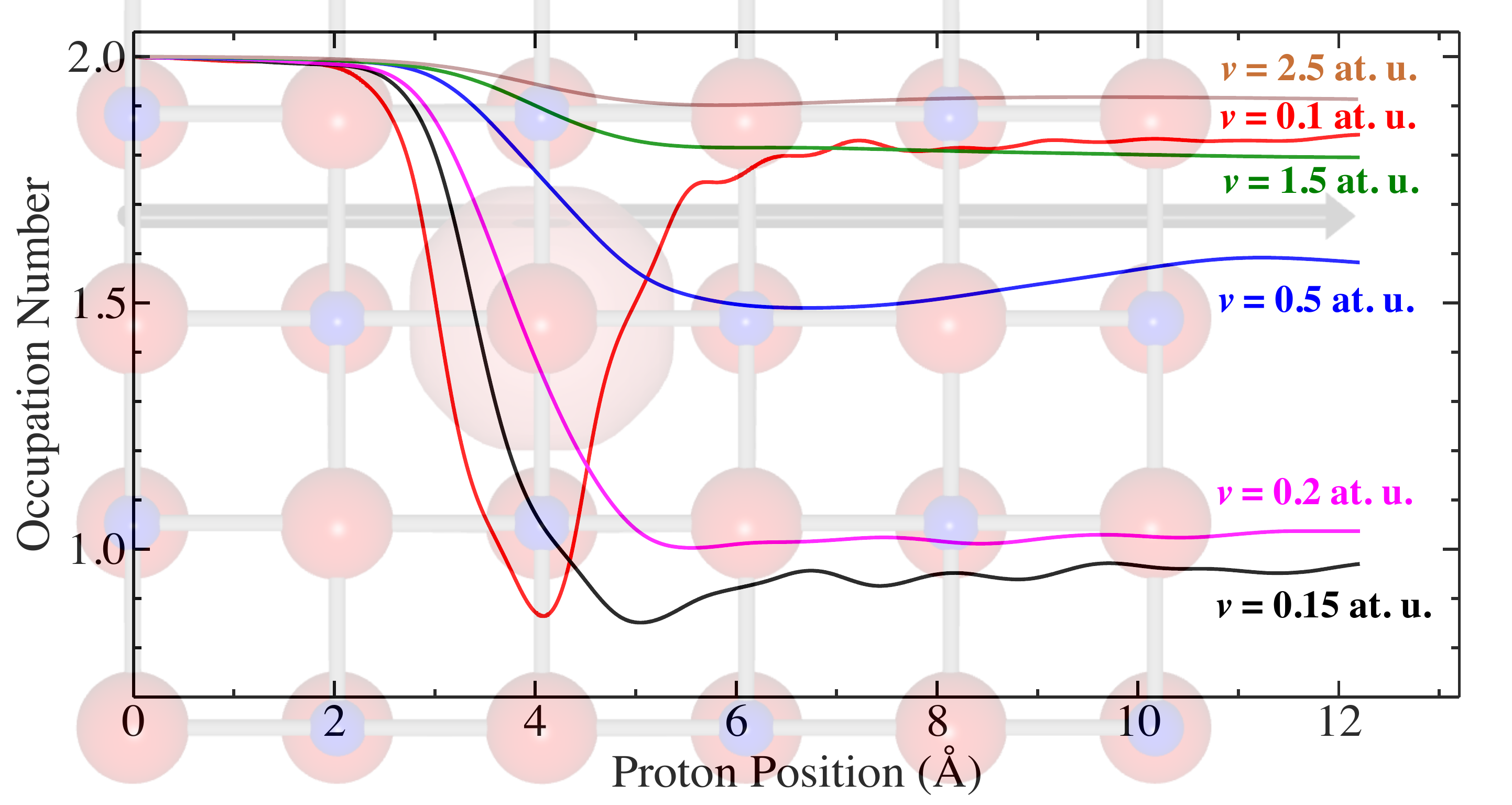}
\caption{\label{fig:occ_v_dep}(Color online.)
Time evolution of the occupation number of the oxygen-vacancy mid-gap defect state during proton irradiation.
The black arrow indicates the projectile trajectory.
Red isosurfaces indicate partial charge density of the mid-gap state at 50\,\% of the maximum value.
}
\end{figure}

We analyze these electronic excitations in detail using time-dependent occupation numbers, Eq.\ \eqref{eq:occ}.
In Fig.\ \ref{fig:change_occ}, we visualize these after the proton with $v$=0.15 at.\ u.\ passes through the simulation cell and is back close to its starting point.
This shows that the aforementioned maximum \emph{increase} of electronic stopping due to the presence of a vacancy is attributed to excitations of almost one (0.97) electron from the vacancy-related mid-gap level, while valence states remain largely unaffected.
Figure \ref{fig:change_occ} also shows that the majority of this excitation, about 0.72 electrons, occupies localized, vacancy-related conduction states (arrows in Fig.\ \ref{fig:change_occ}).
The remaining weight corresponds to excitations into higher-energy states.

Next, we focus on ultrafast electron dynamics and evolution of the vacancy charge state, following proton irradiation.
We illustrate in Fig.\ \ref{fig:occ_v_dep} that the time-dependent number of electrons excited out of the mid-gap level, which shows the largest change of occupation, is strongly dependent on proton velocity \footnote{For convenience, the atomic configuration at $t$=0 was used as reference state for Fig.\ \ref{fig:occ_v_dep}. However, we emphasize that the difference compared to using the atomic configuration at time $t$ is insignificant, since the atomic movement is very small on this time scale.}.
The largest number of electrons is excited for $v$=0.15 at.\ u., coinciding with the velocity for which electronic stopping is enhanced most compared to ideal bulk.
This directly ties the maximum stopping enhancement to a maximum depopulation of the vacancy-related mid-gap level.

In the following, we show that this localized excitation of vacancy-level electrons significantly impacts ion dynamics.
To this end, Fig.\ \ref{fig:Mg2_111_disp} compares the displacement that results from EMD and Born-Oppenheimer MD, for one Mg atom (indicated by blue arrow in the inset).
While Fig.\ \ref{fig:Mg2_111_disp} only shows the upward displacement of one specific Mg nearest neighbor of the initially neutral oxygen vacancy,
in the supplemental material at [URL] we provide a similar analysis for the mean-squared displacement of all first and second nearest neighbors, leading to similar conclusions.

Our analysis indicates that, depending on the projectile kinetic energy, the displacement is enhanced by up to one order of magnitude in the presence of electronic excitations, compared to ground-state Born-Oppenheimer MD.
Figure \ref{fig:Mg2_111_disp} also shows that the vibrational period increases by a factor of up to 2.2 in the presence of excited electrons and that EMD predicts the largest displacement of the nearest-neighbor Mg atom for protons with $v$=0.15 at.\ u.
This is consistent with our earlier discussion that maximum depopulation of the vacancy-defect level (see Fig.\ \ref{fig:occ_v_dep}) and maximum enhancement of electronic stopping occur for that same velocity and, hence, are related to the defect state.
We note that this maximum Mg displacement, induced by radiation of a certain kinetic energy ($v$=0.15 at.\ u.), corresponds to an effective maximum opening of the diffusion path for oxygen, shown in the inset of Fig.\ \ref{fig:Mg2_111_disp}.
Born-Oppenheimer MD qualitatively differs by finding much smaller maximum displacement that occurs for the lowest projectile velocity (see Fig.\ \ref{fig:Mg2_111_disp}), since this corresponds to the longest interaction time of the proton with the Mg atom.
This clearly highlights the intimate coupling of electronic excitations and ionic motion.

To understand this further, we note that the vacancy-related mid-gap level is localized, as can be seen in the insets of Fig.\ \ref{fig:change_occ}, and occupied with two bonding electrons in its ground state \cite{Rinke:2012}.
The excitation of electrons from this state into higher-energy vacancy states, the corresponding displacement of nearest-neighbor Mg atoms, and the increased oscillation period can be understood within the bond-softening model \cite{Itoh:2009}:
The removal of electrons from the bonding state causes bond weakening, or softening, and, thus, a displacement of atoms near the vacancy away from it.
Our EMD simulations quantitatively predict the underlying electron dynamics and the resulting motion of ions.
While longer EMD simulations, coupled to a heat bath, would be desirable to understand atomic diffusion in the presence of electronic excitations, these are prohibitively expensive.

Instead, to account for the statistical character of diffusion, we bridge the gap to the short time scale of EMD using transition-state theory \cite{Eyring:1935}.
More specifically, we quantify hot-electron mediated ion diffusion using the equation for atomic diffusivity $D$ \cite{Phillips:2001}.
This accounts for the probability of a defect to exist, possible routes to diffuse, and the probability to diffuse.
We only consider dynamics at short time scales, which justifies the assumption of constant equilibrium concentration of oxygen vacancies.
Hence, we focus only on the last term, i.e.\ the successful-jump frequency $\Gamma$ of a diffusion event, described by the Vineyard expression \cite{Vineyard:1957},
\begin{equation}
\label{eq:Gamma}
\Gamma = \nu^{*}\exp{\left(-\Delta E_\mathrm{m}/k_\mathrm{B}T\right)},
\end{equation}
with $\nu^{*}$ as attempt (Einstein) frequency and $\Delta E_\mathrm{m}$ as migration barrier.
Using a finite-difference method and  displacements of 0.01 \AA\ we find results for $\nu^{*}$ comparable to 13 THz reported in the literature for oxygen atoms in MgO \cite{Runevall:2011} (see supplemental material at [URL] for explicit results for $\nu^{*}$).
We evaluate $\Delta E_\mathrm{m}$ via the commonly used climbing-image NEB method  \cite{Henkelman:2000_a,Henkelman:2000_b}, as implemented based on the VASP code.

Most importantly, we use constrained DFT (c-DFT) to incorporate occupation-number constraints that account for hot electrons, when computing $\nu^{*}$ and $\Delta E_\mathrm{m}$ using finite-difference and NEB calculations, respectively.
Here we compare three different approximations for the constraint:
(i) KS occupation numbers from EMD using Eq.\ \eqref{eq:occ} represent the most accurate reference immediately after the excitation.
(ii) Alternatively, motivated by previous studies \cite{Lei:2015,Mulroue:2011}, we use an approach that assumes that only the vacancy charge state changes as consequence of proton irradiation.
We use the occupation number of the mid-gap state from our EMD simulation, which is 0.97, as reported above.
(iii) For additional comparison, we model hot, fully \emph{thermalized} electrons within Mermin DFT and an effective Fermi temperature \cite{Mermin:1965,Alavi:1994,Silvestrelli:1996}.
To this end, we compute the total-energy change upon excitation by the proton (see supplemental material at [URL] for detailed description of this procedure).
Neglecting entropy differences, we estimate the Fermi temperature as the temperature that leads to the same total-energy difference and find $T$=9211 K for $v$=0.15 at.\ u.

From this analysis, we find that all three occupation constraints, which we use to mimic hot-electron distributions, give rise to enhanced atomic diffusion and lower migration barriers $\Delta E_\mathrm{m}$ (see inset of Fig.\ \ref{fig:Mg2_111_disp}) compared to the ground-state.
The values of $\Delta E_\mathrm{m}$ are 0.34, 1.07, and 1.33 eV lower than the ground-state barrier of $\approx$ 4.4 eV when Mermin DFT, fixed-charge model, and c-DFT are used, respectively.
The difference between c-DFT and Mermin DFT is related to the underlying time scale:
c-DFT is the better approximation at early stages after proton irradiation, well before thermalized excited electrons dominate over non-thermalized ones.
While this is a heavily debated question, early tests (see supplemental material at [URL] for discussion of the long-term evolution of occupation numbers) 
indicate that this is the case for the first several tens of femtoseconds after irradiation, during which Mermin DFT is not adequate.
The fixed-charge model predicts results close to c-DFT, highlighting that not only the charge state of the point defect but also the excited-electron distribution enhance atomic diffusion.

The influence of hot electrons on phonon frequency and, thus, attempt frequency $\nu^{*}$ is more complicated:
We find slightly enhanced (+1.10\,\%), significantly enhanced (+19.85\,\%), and reduced ($-6.61$\,\%) attempt frequencies within fixed-charge model, Mermin DFT, and c-DFT, respectively.
However, while excited-electron distributions affect migration barrier and attempt frequency, overall the migration barrier dominates the resulting diffusivity.
Equation \eqref{eq:Gamma} unveils an exponential dependence of $\Gamma$ on $\Delta E_\mathrm{m}$.
The 20\,\% change of the effective jump rate is much smaller compared to the change in diffusivity.
Using $T$=900 K, the difference in jump frequency between the migration barrier calculated using ground-state DFT ($\approx 4.4$ eV) and c-DFT ($\approx 3.1$ eV) is on the order of $10^{7}$.
We estimate that, under typical proton irradiation conditions in focused ion beams and vacancy concentrations of at least $\approx 10^{13}$ cm$^{-3}$, enough vacancies are ionized for the additional diffusion contribution to be observable (see supplemental material at [URL] for details).

In summary, we devise a first-principles simulation framework to quantitatively study hot-electron mediated ion diffusion, by combining real-time time-dependent density functional theory, occupation-number constraints, and the nudged-elastic band method.
Our parameter-free technique bridges time scales ranging from ultrafast electronic-excitation dynamics to ion diffusion across migration barriers of several eV.
We apply this framework to magnesium oxide with and without an oxygen vacancy.
Using our results, we discover a novel diffusion mechanism that derives from a significant lowering of migration energies in the presence of non-thermalized or thermalized carriers.
We observe a strong dependence of this mechanism on the projectile kinetic energy and attribute this to excitations of specific defect electrons.
This implies that hot-electron mediated ion diffusion is defect-specific and, for each defect, occurs only at a specific penetration depth of the projectile in the target, where it can efficiently excite defect electrons.
Furthermore, our findings illustrate a possible route towards deliberate diffusion enhancement by irradiating with projectiles of a specific kinetic energy.
We envision that this can be used to actively enhance or suppress defect diffusion by tuning the energy of the ion beam.

\begin{acknowledgments}
Fruitful discussions with Ravi Agarwal, Xavier Andrade, Alfredo Correa, Yosuke Kanai, and Pascal Pochet are gratefully acknowledged.
Financial support from the Sandia National Laboratory-UIUC collaboration is acknowledged (SNL grant no.\ 1736375).
C.-W.\ L.\ acknowledges support from the Government Scholarship to Study Abroad from the Taiwan Ministry of Education.
An award of computer time was provided by the Innovative and Novel Computational Impact on Theory and Experiment (INCITE) program.
This research used resources of the Argonne Leadership Computing Facility, which is a DOE Office of Science User Facility supported under Contract DE-AC02-06CH11357.
Data used in this work are available at the Materials Data Facility \cite{MDF,data}.
\end{acknowledgments}

\bibliography{MgO_paper.bib}

\end{document}